%% file: main_arxiv.tex
\documentclass{article}
\usepackage{ijcai18}

\usepackage{times}
\usepackage{xcolor}
\usepackage{soul}
\usepackage[utf8]{inputenc}
\usepackage[small]{caption}

\usepackage{enumerate,paralist}
\usepackage{parskip}
\usepackage{multirow}
\usepackage{graphicx}
\usepackage{graphics}
\usepackage{color}
\usepackage[T1]{fontenc}
\usepackage{textcomp}
\usepackage{listing}
\usepackage{listings}
\usepackage{tikz,pgffor}
\usetikzlibrary{arrows}
\usetikzlibrary{shapes}
\usepackage{mathtools}

\usetikzlibrary{calc}
\usetikzlibrary{automata}
\usetikzlibrary{positioning}

\tikzstyle{proglabel}=[shape=circle,draw,inner sep=0pt,minimum size=5mm]
\tikzstyle{tran}=[draw,->,>=stealth, rounded corners]

\usepackage{amsmath}
\usepackage{amssymb}
\usepackage{amsthm}

\usepackage{stmaryrd}
\usepackage{url}                  

\usepackage{listings}
\lstdefinelanguage{prog}
{
morekeywords={prob, if, then, else, fi, while, do, od, true, false, and, or, skip},
sensitive = false
}

\newcommand{\Rset}{\mathbb{R}}
\newcommand{\Nset}{\mathbb{N}}

\newcommand{\probm}{\mathbb{P}}
\newcommand{\expv}{\mathbb{E}}
\newcommand{\pv}{\mathbf{v}}
\newcommand{\rv}{\mathbf{u}}

\newcommand{\condexpv}[2]{{\expv}{\left({#1}{\mid}{#2}\right)}}

\newtheorem{definition}{Definition}

\newtheorem{theorem}{Theorem}
\newtheorem{example}{Example}
\newtheorem{remark}{Remark}

\title{Computational Approaches for Stochastic Shortest Path on Succinct MDPs
\thanks{The research was partially supported by Vienna Science
	and Technology Fund (WWTF) Project ICT15-003, Austrian Science Fund
	(FWF) NFN Grant No S11407-N23 (RiSE/SHiNE), and ERC Starting grant
	(279307: Graph Games).}
}

\author{
Krishnendu {Chatterjee}{$^1$},
Hongfei {Fu}{$^2$},
Amir Kafshdar {Goharshady}{$^1$}, and
Nastaran {Okati}{$^3$}
\\
${^1}$ IST Austria (Institute of Science and Technology Austria) \\
${^2}$ Shanghai Jiao Tong University\\
${^3}$ Ferdowsi University of Mashhad\\
kchatterjee@ist.ac.at,
fuhf@cs.sjtu.edu.cn,
goharshady@ist.ac.at,
nastaran.okati@mail.um.ac.ir
}

\usepackage{tabularx}
\usepackage{syntax} 

\newcolumntype{L}[1]{>{\raggedright\arraybackslash}p{#1}}
\newcolumntype{C}[1]{>{\centering\arraybackslash}p{#1}}
\newcolumntype{R}[1]{>{\raggedleft\arraybackslash}p{#1}}

\begin{document}

\maketitle

\input{abstract\string_arxiv}

\input{introduction\string_arxiv}

\input{language\string_arxiv}

\input{cost\string_arxiv}

\input{algorithm\string_arxiv}

\input{cases\string_arxiv}

\input{experiments\string_arxiv}

\input{related\string_arxiv}

\input{conclusion\string_arxiv}



\bibliographystyle{named}
\bibliography{PL}


\end{document}

%% file: abstract_arxiv.tex
\begin{abstract}
We consider the stochastic shortest path (SSP) problem
for succinct Markov decision processes (MDPs), where the MDP consists of
a set of variables, and a set of nondeterministic rules that update the
variables.
First, we show that several examples from the AI literature 
can be modeled as succinct MDPs.
Then we present computational approaches for upper and lower bounds for the SSP problem:
(a)~for computing upper bounds, our method is polynomial-time in the
implicit description of the MDP; 
(b)~for lower bounds, we present a polynomial-time (in the size
of the implicit description) reduction to quadratic programming.
Our approach is applicable even to infinite-state MDPs.
Finally, we present experimental results to demonstrate the effectiveness of our approach
on several classical examples from the AI literature.
\end{abstract}

%% file: introduction_arxiv.tex
\section{Introduction}

\noindent{\em Markov decision processes.}
Markov decision processes (MDPs)~\cite{Howard} are a standard mathematical model for
sequential decision making, with a wide range of applications in
artificial intelligence and beyond~\cite{Puterman:1994:MDP:528623,VriezeMDP,DBLP:books/lib/Bertsekas05}.
An MDP consists of a set of states, a finite set of actions (that
represent the nondeterministic choices), and a probabilistic transition
function that describes the transition probability over the next states,
given the current state and action.
One of the most classical optimization objectives in MDPs is the
stochastic shortest path (SSP) problem, where the transitions of
the MDP are labeled with rewards/costs, and the goal is to optimize
the expected total rewards until a target set is reached.

\noindent{\em Curse of dimensionality.} In many typical
applications, the computational analysis of MDPs suffers from the
curse of dimensionality.
The state space of the MDP is huge as it represents valuations to
many variables that constitute the MDP.
A well-studied approach for algorithmic analysis of large MDPs is
to consider factored MDPs~\cite{DBLP:journals/jair/GuestrinKPV03,DBLP:journals/ai/DelgadoSB11}, where the transition and reward
function dependency can be factored only on few variables.

\noindent{\em Succinct MDPs.} In the spirit of factored
MDPs, which aim to deal with the curse of dimensionality, we consider
{\em succinct MDPs}, where the MDP is described implicitly by a set of
variables, and a set of rules that describe how the variables are updated.
The rules can be chosen non-deterministically from this set at every time step
to update the variables, until a target set (of valuations) is reached.
The rules and the target set represent the succinct description of the
MDP. We consider the SSP problem for succinct MDPs.

\noindent{\em Our contributions.} Our main contributions are as follows:
\begin{compactenum}
\item First, we show that many examples from the AI literature (e.g.,
Gambler's Ruin, Robot Planning, and variants of Roulette)
can be naturally modeled as succinct MDPs.

\item Second, we present mathematical and computational results for the SSP problem
for succinct MDPs.
For the SSP problem the $\sup$-value (resp. $\inf$-value) represents the expected shortest
path value with supremum (resp., infimum) over all policies.
We consider linear bounds for the SSP problem and our algorithmic bounds are as follows:
(a)~for the $\sup$-value (resp. $\inf$-value) we show that an upper (resp., lower) bound
can be computed in polynomial time in the implicit description of the MDP; 
(b)~for the $\sup$-value (resp. $\inf$-value) we show that a lower (resp., upper)
bound can be computed by a polynomial-time (in the implicit description) reduction to
quadratic programming.
Our approach is as follows: we use results from probability theory to establish
a mathematical approach to compute bounds for the SSP problem for succinct MDPs (Section~\ref{sect:theory}),
and reduce the mathematical approach to constraint solving to obtain
a computational method (Section~\ref{sect:algorithm}).

\item Finally, we present experimental results on several classical examples from
the literature where our method computes tight bounds (i.e., lower and upper bounds are
quite close) on the SSP problem extremely efficiently.
\end{compactenum}

\noindent{\em Comparison with approaches for factored MDPs.}
Some key advantages of our approach are the following.
First, our approach gives a provably polynomial-time algorithm or polynomial-time
reduction to quadratic programming in terms of the size of the implicit description of the MDP.
Second, while algorithms for factored MDPs are typically suitable for finite-state
MDPs, our method can handle MDPs with countable state space (e.g., when the
domains of the variables are integers), or even uncountable state space (e.g.,
when the domains of the variables are reals). See Remark~\ref{remark:fmdp} for further details.

%% file: language_arxiv.tex
\newcommand{\AS}{Q}
 \newcommand{\ndc}{\ell}

\section{Definitions and Examples}

We define succinct Markov decision processes,
the stochastic shortest path problem, and provide illustrative examples.

\subsection{Succinct Markov Decision Processes}

\noindent{\em Markov decision processes (MDPs).}
MDPs are a standard mathematical model for sequential decision
making.
An MDP consists of a state space $S$ and a finite action space $A$,
and given a state $s$ and an action $a$ permissible in $s$, there is a probability
distribution function $P$ that describes the transition probability
from the current state to the next states (i.e., $P(s' \mid s,a)$ is the transition
probability from $s$ to $s'$ given action $a$).
Moreover, there is a reward assigned to each state and action pair.

\smallskip\noindent{\em Factored MDPs.}
In many applications of MDPs, the state space of the MDP is high-dimensional and
huge (i.e., the state space consists of valuations to many variables).
For computational analysis it is often considered that the MDP can be factored,
i.e., the transition and reward probabilities depend only on a small number of
variables. For algorithmic analysis of finite-state factored MDPs see~\cite{DBLP:journals/jair/GuestrinKPV03}.

\smallskip\noindent{\em Succinct MDPs.}
In this work, we consider succinct MDPs, which are related to factored MDPs.
First, we present an informal description and then the details.
A {\em succint MDP} is described by a set of variables, and a set of rules
that update them. The update rule can be
chosen non-deterministically or stochastically.
Thus the MDP is described implicitly with the set of rules and a condition
that describes the target set of states, and the MDP terminates when the
target set is reached.
We describe succinct MDPs as a special class of programs with a 
single while loop.

\smallskip\noindent{\em Simple-while-loop succinct MDPs.}
We consider succinct MDPs described by a simple while loop program.
There are two types of variables, namely, \emph{program} and \emph{sampling} variables.
Program variables are normal variables, while sampling variables are those whose values
are sampled independently  wrt some probability distribution.
In general, both program and sampling variables can take integer, or even real values.
The succinct MDP is described by a simple while loop of the form:
\begin{equation}\label{eq:swl}
\textbf{while}~\phi~\textbf{do}~\AS_1\square\dots \square \AS_k~\textbf{od}
\end{equation}
\begin{compactitem}
\item $\phi$ is the \emph{loop guard} defined as a single comparison between
linear arithemtic expressions over program variables (e.g. $x\le y+1$);
\item in the loop body, $\AS_1,\dots, \AS_k$ are sequential compositions of assignment statements,
grouped by the non-determinism operator $\Box$.
\end{compactitem}
Every assignment statement has a program variable as its left-hand-side and a linear arithmetic expression over program and sampling variables as its right-hand-side.
The operator $\Box$ is the nondeterministic choice which means that the decision as to which $\AS_i$ will be executed in the current loop iteration depends on a scheduler (or policy)
that resolves nondeterminism.
We first provide the formal syntax and then a simple example.

\smallskip\noindent{\em Formal syntax of simple-while-loop programs.}
A succinct MDP is specified by a simple-while-loop program equipped with probability distributions for sampling variables. We now formalize the intuitive description provided in Equation~(\ref{eq:swl}).
Formally, a simple-while-loop program can be produced using the following grammar, where each $\texttt{<pvar>}$ is chosen from a finite fixed set $X$ of program variables, each $\texttt{<svar>}$ from a finite fixed set $R$ of sampling variables and each $\texttt{<constant>}$ denotes a floating point number:

\begin{grammar}
	<simple-while-loop-program> ::= `while' <guard> `do' `{' <nondet-block-list> `}' `od'
	
	<guard> ::= <linear-pvar-expr> <cmp> <linear-pvar-expr>
	
	<linear-pvar-expr> := <constant> | <constant> `*' <pvar>
	\alt <linear-pvar-expr> `+' <linear-pvar-expr>
	
	<cmp> ::= `>=' | `>' | `<=' | `<'
	
	<nondet-block-list> ::= <block> 
	\alt <block> $\square$ <nondet-block-list>
	
	<block> ::= <assignment> | <assignment> <block>
	
	<assignment> ::= <pvar> `:=' <linear-expr> `;'
	
	<linear-expr> ::= <constant> | <constant> `*' <pvar>
	\alt <constant> `*' <svar>
	\alt <linear-expr> `+' <linear-expr>
	
\end{grammar}

\smallskip
\begin{example}\label{ex:running}
Consider the following program
\[
\mbox{\textbf{\em while }}x\ge 1~\textbf{\em do } x:=x+r~\square~ x:=x-1~\textbf{\em od}
\]
where $x$ is a program variable and $r$ is a sampling variable that observes the two-point distribution
$\probm(r={-1})=\probm(r={1})=\frac{1}{2}$.
Informally, the program performs either decrement or random increment/decrement on $x$ until its value is zero.
\end{example}

\noindent{\em Informal description of the semantics.}
Given a simple-while-loop program in the form (\ref{eq:swl}), an MDP is obtained as follows:
the state space $S$ consists of values assigned to program variables (i.e., \emph{valuations} for program variables);
the action space $A$ correponds to the non-deterministic choice between $\AS_1,\ldots,\AS_k$;
and the transition function $P$ depends on the sampling of the sampling variables, which given
the current valuation for program variables probabilistically updates the valuation.
The assignments are linear functions, and the loop guard $\phi$ describes the target
states as those which do not satisfy $\phi$.
Given the above components, the notion of a policy (or scheduler) $\sigma$
that resolves the non-deterministic choices,  and that of the probability space
$\probm^{\sigma}$ given a policy is standard~\cite{Puterman:1994:MDP:528623}. For a more formal treatment of the semantics, see Section~\ref{sec:formal}.

\smallskip
\begin{remark}[Simple-while-loop MDPs and Factored MDPs] \label{remark:fmdp}
The principle behind factored MDPs and simple-while-loop succinct MDPs is similar.
Both aim to consider high-dimensional and large MDPs described implicitly in a succinct way.
In factored MDPs the transitions and reward functions can be factored based on small sets of variables, 
but the dependency on the variables in these sets can be arbitrary. 
In contrast, in simple-while-loop succinct MDPs, we only allow linear arithmetic expressions as guards and assignments. 
Moreover, our  MDPs do not allow nesting of while loops. 
The goal of our work is to consider linear upper and lower bounds, and nesting of linear loops can result in super-linear bounds. hence we do not consider nesting of loops. 
Therefore, simple-while-loop succinct MDPs are a special class of factored MDPs.

For algorithmic approaches with theoretical guarantees on computational complexity, 
the analysis of factored MDPs has typically been restricted to finite-state MDPs. 
However, we will present solutions for simple-while-loop programs, where the variables can take integer or real values,
and thus the underlying state space is infinite or even uncountable.
Thus the class of simple-while-loop MDPs consists of large finite-state MDPs (when the variables are bounded);
countable state MDPs (variables are integer); and even uncountable state MDPs (variables are real-valued). 
Moreover, our algorithmic approaches provides computational complexity guarantee on the input size of the implicit representation 
of the MDP.
Note that for finite-state MDPs, the implicit representation can be exponentially more compact than the explicit MDP. 
For example, $n$ boolean variables lead to a state-space of size $2^n$. 
\end{remark}

In the sequel we consider MDPs obtained from simple-while-loop programs and, for brevity, call them succinct MDPs.


%% file: cost_arxiv.tex
\newcommand{\While}{\textbf{while}}
\newcommand{\infval}{\mbox{\sl infval}}
\newcommand{\supval}{\mbox{\sl supval}}
\newcommand{\Do}{\textbf{do}}
\newcommand{\Od}{\textbf{od}}
\newcommand{\If}{\textbf{if}}
\newcommand{\Else}{\textbf{else}}
\lstdefinelanguage{ours}
{
	morekeywords={
		if,
		while,
		do,
		od,
		else,
		?,
		reward,
		cost
	},
	sensitive=false, 
	morecomment=[l]{//}, 
	morecomment=[s]{/*}{*/}, 
	morestring=[b]" 
}
\lstset{literate={?}{{$\square$}}1 {ã}{{\~a}}1 {é}{{\'e}}1,
}
\lstset{language=ours}

\subsection{Stochastic Shortest Path on Succinct MDPs}

We consider the stochastic shortest path (SSP) problem on succinct MDPs.
Below we fix a succinct MDP described by a while-loop in the form~(\ref{eq:swl}).

\noindent{\em Reward function.}
We consider a reward function $R$ that assigns a reward $R(\mathbf{u},\ell)$ when
the sampling valuation for sampling variables is $\mathbf{u}$ and the non-deterministic choice
is $\AS_\ell$ (in a loop iteration).
We assume that there is a maximal constant $R_{\max}\ge 0$ such that $|R(\mathbf{u},\ell)|\le R_{\max}$ for all $\rv,\ell$.
The rewards need not be nonnegative as our approach is able to handle negative rewards as well.

\noindent{\em Stochastic shortest path.}
Given an initial valuation $\pv$ for program variables and a policy $\sigma$,
the definition of expected total reward/cost until termination is standard.
The $\inf$-value (resp., $\sup$-value) of a succinct MDP, given an initial valuation $\pv$
for program variables, is the infimum (resp., supremum) expected reward value over all policies
that ensure finite expected termination time, which we denote as
$\infval(\pv)$ (resp., $\supval(\pv)$).

\noindent{\em Computational problem.}
We consider the computational problem of obtaining upper and lower bounds
for the $\inf$-value and $\sup$-value for succinct MDPs.
Due to the similarity of the problems, we will focus only on computing lower and
upper bounds for the $\sup$-value, and the results for $\inf$-value are similar and
omitted.

\subsection{Illustrative Examples}\label{sect:runningexamples}

\begin{example}[Gambler's Ruin] \label{examp:gr}
We start with a simple and classical example. A gambler has $x_0$ tokens for gambling.
In each round he can choose one of the two types of gambling.
In type 1, he wins with probability $p_1 < 1/2$ and in type 2 with probability $p_2 < 1/2$.
A win leads to a reward of $w$ and an extra token.
A loss costs one token. The player gambles as long as he has tokens left.
His goal is to maximize overall expected reward. Letting $p_1 = 0.4, p_2=0.3$ and $w = 1$, a succinct MDP for this example is shown in Figure \ref{fig:gambler}.
\begin{figure}[!hb]
\begin{lstlisting}[mathescape]
while $x \ge 1$ do
  ? if($0.4$) { $x := x + 1$  reward=$1$ }
    else   { $x := x - 1$ }

  ? if($0.3$) { $x := x + 1$  reward=$1$ }
    else   { $x := x - 1$ }
\end{lstlisting}

\caption{Gambler's Ruin as a succinct MDP}
\label{fig:gambler}

\end{figure}
\end{example}




\smallskip
\begin{remark}
Note that above we use probabilistic \textbf{if} as syntactic sugar,
where the assignments in \textbf{if}$(p)$ run with probability $p$ and those in the \textbf{else} part with probability $1-p$.
Given that the assignments are linear, this can be translated back to a succinct MDP, e.g. the first \textbf{if}-\textbf{else} block in Figure \ref{fig:gambler} is equivalent to:

\begin{lstlisting}[mathescape]
	$x := x + r$  reward=$0.4$
\end{lstlisting}

where $\textup{r}$ is a sampling variable with $\probm(r = 1) = 0.4$ and $\probm(r = -1) = 0.6$.
\end{remark}

\smallskip
\begin{example}[Continuous variant] \label{examp:grcts}
We can also consider a continuous variant of the example where the sampling variable
$r$ is chosen from some continuous distribution with expected value~$-0.2$
(e.g., uniform distribution $[-0.8,0.4]$). 
\end{example}

\subsection{Technical Details of Semantics and SSP} \label{sec:formal}
\subsubsection{Formal Semantics}
We formalize our semantics by introducing the notion of valuations which specify current values for program and sampling variables.
Below we fix a program $P$ in the form (\ref{eq:swl}) and let $X$ (resp. $R$) be the set of program (resp. sampling) variables appearing in $P$. The sizes of $X$ and $R$ are denoted by $|X|$ and $|R|$, respectively.
We also impose arbitrary linear orders on both of $X$ and $R$, and assume that for each sampling variable $r_i \in R$, a distribution $\delta_i$ is given and each time $r_i$ appears in the program, its value is stochastically sampled according to $\delta_i$.

\smallskip
{\em Valuations.} A \emph{program valuation} is a vector $\pv\in\Rset^{|X|}$.
Intuitively, a valuation $\pv$ specifies that for each $x\in X$ with rank $i$ in the linear order, the value assigned is the $i$th coordinate $\pv[i]$ of $\pv$.
Likewise, a \emph{sampling valuation} is a vector $\rv\in\Rset^{|R|}$.



For each program valuation $\pv$, we say that $\pv$ \emph{satisfies} the loop guard $\phi$, denoted by $\pv\models\phi$ , if the formula $\phi$ holds when every appearance of a program variable is replaced by its corresponding value in $\pv$. Moreover, each $Q_m$ in $P$ ($1\le m\le k$) now encodes a function $F_m: \Rset^{|X|}\times \Rset^{|R|}\rightarrow  \Rset^{|X|}$ which transforms the program valuation $\pv$ before the execution of $Q_m$ and the sampled values in $\rv$ into the program valuation $F_m(\pv, \rv)$. By our linear setting, each $F_m$ is also linear.

Our semantics are based on \emph{paths}. Intuitively, a path is an infinite sequence of valuations
where the valuations at the $n$th step reflects the current values for program and sampling variables at the $n$th step.

{\em Paths.} A \emph{path} is a finite or infinite sequence of triples $\{(\pv_n,\rv_n, \ndc_n)\}_{n\ge 0}$ such that each $\pv_n$ (resp. $\rv_n, \ndc_n$) is a program valuation (resp. sampling valuation, nondeterministic choice). 
The intuition is that each $\pv_n$ (resp. $\rv_n$) is the current program valuation (resp. sampling valuation, nondeterministic choice)
right before the $n$th loop-iteration of $P$.

The program $P$ may involve nondeterministic choices (i.e., the operator $\Box$) which are still unspecified.
To resolve nondeterminism, we need the standard notion of schedulers.

\noindent{\em Schedulers.}
A \emph{scheduler} is a function which maps each finite path $(\pv_0,\rv_0,\ndc_0),\dots,(\pv_{n-1},\rv_{n-1},\ndc_n)$ and current program valuation $\pv_{n}$ to a number $m$ in $\{1,2,\dots,k\}$ representing
the choice of $Q_m$ ($1\le m\le k$) at the $n$th loop iteration.

Intuitively, a scheduler resolves the nondeterministic choice at each iteration of $P$ by choosing which $Q_m$ to run in the loop body.
The resolution at the $n$th iteration may depend on all previous valuations $P$ has traversed before.

\emph{Intuitive Semantics.} Consider an initial program valuation $\pv$ and a scheduler $\sigma$. An infinite path $\{(\pv_n,\rv_n,\ndc_n)\}_{n\in\Nset_0}$ is constructed as follows.
Initially, $\pv_0=\pv$. Then at each step $n$ ($n\ge 0$):
first, a sampling valuation $\rv_{n}$ is obtained through samplings for all sampling variables, where the sampling of each sampling variable observes a predefined probability distribution for the variable;
second, if $\pv_n\models\phi$ then the program enters the loop and
$\ndc_n:=\sigma(\{(\pv_l,\rv_l)\}_{1\le l\le n-1}, \pv_n)$,
$\pv_{n+1}:=F_{\ndc_n}(\pv_n,\rv_n)$, otherwise the program terminates and $\pv_{n+1}:=\pv_n$.

{\em Probability Space.} A \emph{probability space} is a triple $(\Omega,\mathcal{F},\probm)$, where $\Omega$ is a nonempty set (so-called \emph{sample space}), $\mathcal{F}$ is a \emph{$\sigma$-algebra} over $\Omega$ (i.e., a collection of subsets of $\Omega$ that contains the empty set $\emptyset$ and is closed under complementation and countable union), and $\probm$ is a \emph{probability measure} on $\mathcal{F}$, i.e., a function $\probm\colon \mathcal{F}\rightarrow[0,1]$ such that (i) $\probm(\Omega)=1$ and
(ii) for all set-sequences $A_1,A_2,\dots \in \mathcal{F}$ that are pairwise-disjoint
(i.e., $A_i \cap A_j = \emptyset$ whenever $i\ne j$)
it holds that $\sum_{i=1}^{\infty}\probm(A_i)=\probm\left(\bigcup_{i=1}^{\infty} A_i\right)$.
Elements $A\in\mathcal{F}$ are usually called \emph{events}.
An event $A\in\mathcal{F}$ is said to hold \emph{almost surely} (a.s.) if $\probm(A)=1$.

\emph{Formal Semantics.} Given an initial program valuation $\pv$ and a scheduler $\sigma$,
We build the probability space $(\Omega,\mathcal{F},\probm^{\sigma}_{\pv})$ for the program $P$ as follows.
First, $\Omega$ is the set of all infinite paths.
Then, we construct $\mathcal{F},\probm^{\sigma}_{\pv}$ through general state-space Markov chains.
In detail, we build the kernel function on the set of all finite paths.
The construction of the kernel function follows exactly from our aformentioned intuitive semantics.
Then the probability space on infinite paths is generated uniquely from the kernel function.

\subsubsection{Formal Details of the SSP Problem}
We consider accumulated cost until program termination.
We first define several classic notions of probability theory.

\noindent{\em Random Variables.} A \emph{random variable} $X$ for a probability space $(\Omega,\mathcal{F},\probm)$
is an $\mathcal{F}$-measurable function $X\colon \Omega \rightarrow \Rset \cup \{-\infty,+\infty\}$, i.e.,
a function satisfying the condition that for all $d\in \Rset \cup \{-\infty,+\infty\}$, the set $\{\omega\in \Omega\mid X(\omega)<d\}$ belongs to $\mathcal{F}$;
By convention, we abbreviate $+\infty$ as $\infty$.

Below we fix a program $P$ in the form (\ref{eq:swl}) with program variables $X$ and sampling variables $R$.
We first establish the notion of cost functions for $P$ which measures the cost consumed in one loop iteration.
In this paper, we consider that the cost is only related to the sampling valuation and the nondeterministic choice before the execution of the loop body.

\smallskip
\begin{definition}
	A cost function is a function
	$C:\Rset^{|R|}\times \{1,\dots,k\}\rightarrow [0,\infty)$.
\end{definition}
The next definition introduces the notion of accumulated cost.

\smallskip
\begin{definition}
	For each $m\in\Nset$, we define the random variable $C_m$ for \emph{cost at the $m$th step} by:
	\[
	C_m(\{(\pv_n,\rv_n,\ndc_n)\}_{n\in\Nset}):=
	\begin{cases}
	R(\rv_m, \ndc_m) & \mbox{if }\pv_m\models\phi\\
	0 & \mbox{otherwise}
	\end{cases}
	\]
	for any infinite path $\{(\pv_n,\rv_n,\ndc_n)\}_{n\in\Nset}$.
	Then the random variable $C_\infty$ for \emph{accumulated cost until termination} is defined as
	$C_\infty(\rho):=\sum_{n=0}^\infty C_n(\rho)$ for all infinite paths $\rho$.
\end{definition}

\noindent{\em Expectation.} The \emph{expected value} of a random variable $X$ from a probability space $(\Omega,\mathcal{F},\probm)$, denoted by $\expv(X)$, is defined as the Lebesgue integral of $X$ w.r.t $\probm$, i.e.,
$\expv(X):=\int X\,\mathrm{d}\probm$~;
the precise definition of Lebesgue integral is somewhat technical and is
omitted  here (cf.~\cite[Chapter 5]{probabilitycambridge} for a formal definition).

We study the expected accumulated cost $\expv(C_\infty)$, which is an important criterion for measuring how much cost the program consumes upon termination. In the presence of nondeterminism, we consider maximum and minimum accumulated cost over all schedulers.

\smallskip
\begin{definition}
	The \emph{maximum} (resp. \emph{minimum}) \emph{expected accumulated cost} $\supval(\pv)$ (resp. $\infval(\pv)$) w.r.t an initial program valuation $\pv$
	is defined as $\supval(\pv):=\sup_\sigma \expv^\sigma_\pv(C_\infty)$ (resp. $\infval(\pv):=\inf_\sigma \expv^\sigma_\pv(C_\infty)$), where $\sigma$ ranges over all schedulers and $\expv^\sigma_\pv(\centerdot)$ is the expectation under the probability space $(\Omega,\mathcal{F},\probm^{\sigma}_{\pv})$ for infinite paths.
\end{definition}

\section{Theoretical Results}\label{sect:theory}

In this section we present the main theoretical results which forms the basis of
our algorithms of the following section.

\noindent{\em Notation.} Given a succinct MDP in the form (\ref{eq:swl}), we let $X$ be the set of program variables in the program, $|X|$ be the size of $X$, $\Rset^{|X|}$ be the set of $|X|$-dimensional real-valued vectors,
$\pv\in \Rset^{|X|}$ be a valuation for program variables such that the value for the $i$th program variable is the $i$th coordinate of $\pv$, $\rv$ be a valuation for sampling variables, $N:=\{1,\dots,k\}$ be the set of non-deterministic choices, $\ell\in N$ be a non-deterministic choice, and $F_\ell$ ($\ell\in N$) be the function such that $F_\ell(\pv,\rv)$ is the valuation for program variables resulting from executing $\AS_\ell$ with the valuation $\pv$ for program variables and the sampled valuation $\rv$ for sampling variables. Table~\ref{tab:notations} summarizes the notation.

\begin{table}[!hb]
\centering
\resizebox{\columnwidth}{!}{%

\begin{tabular}{|C{1.3cm}|C{7.4cm}|}
\hline
Notation & Meaning \\
\hline
\hline
$X$ & the set of program variables \\
\hline
$|X|$ & the size of $X$ \\
\hline
$\Rset^{|X|}$ & the set of $|X|$-dimensional real column vectors \\
\hline
$\pv$ & a valuation for program variables \\
\hline
$\rv$ & a valuation for sampling variables \\
\hline
$\infval(\pv)$ & the $\inf$-value for initial valuation $\pv$ \\
\hline
$\supval(\pv)$ & the $\sup$-value for initial valuation $\pv$ \\
\hline
$N$ & the set of non-deterministic choices \\
\hline
$\ell$ & a non-deterministic choice in $N$ \\
\hline
$F_\ell$ & the transformation function of $\AS_\ell$ mapping valuations before its execution to the resulting valuation after it\\
\hline
$R$ & the reward function \\
\hline
$R_{\max}$ & an upperbound for the absolute value of the rewards \\
\hline
\end{tabular}}
\caption{A Summary of Notation}
\label{tab:notations}
\end{table}

\subsubsection{Upper bounds}
We first introduce the main concept for computing an upper bound for $\supval(\pv)$ formally,
and then present the informal description.

\smallskip
\begin{definition}[Linear Upper Potential Functions (LUPFs)]\label{def:lupf}
A \emph{linear upper potential function (LUPF)} is a function $h:\Rset^{|X|} \rightarrow\Rset$
that satisfies the following conditions:
\begin{compactitem}
\item[\emph{(C1)}] $h$ is \emph{linear}, i.e., there exist $\mathbf{a}\in\Rset^{|X|}$ and $b\in\Rset$ such that for all $\pv\in\Rset^{|X|}$, we have $h(\pv)=\mathbf{a}^\mathrm{T}\cdot\pv+b$;
\item[\emph{(C2)}] for all $\pv,\rv,\ell$ such that $\pv\models\phi$, $F_\ell(\pv,\rv)\models\neg\phi$ and $\ell\in N$, we have
$K\le h(F_\ell(\pv,\rv))\le K'$
for some fixed constants $K$ and $K'$;
\item[\emph{(C3)}] for all $\ell\in N$ and all valuations $\pv$ such that $\pv\models\phi$,
\[
h(\pv)\ge \expv_{\rv}(h(F_\ell(\pv,\rv)))+\expv_{\rv}(R(\rv,\ell))
\]
where $\expv_{\rv}(h(F_\ell(\pv,\rv))),\expv_{\rv}(R(\rv,\ell))$ are the expected values over the sampling $\mathbf{u}$ when fixing $\pv$ and $\ell$;
\item[\emph{(C4)}] for all $\pv$ such that $\pv\models\phi$, all sampling valuations $\rv$ and all $\ell\in N$,
we have $|h(\pv)-h(F_\ell(\pv,\rv))|\le M$
for some fixed constant $M\ge 0$.
\end{compactitem}
\end{definition}
Informally, (C1) specifies the linearity of LUPFs, (C2) specifies that the value of the function at terminating valuations should be bounded, (C3) specifies that the current value of the function at $\pv$ is no less than that of the next step at $F_\ell(\pv,\rv)$ plus the cost/reward at the current step, and finally (C4) specifies that the change of value between the current step $\pv$ and the next step $F_\ell(\pv,\rv)$ is bounded.
Note that $\expv_{\rv}(h(F_\ell(\pv,\rv)))$ is linear in $\pv$ since $h$ and $F_\ell$ are linear.
Note that the function $h$ (only) depends on the valuations of the variables before the loop execution,
and hence it is only loop-dependent (but not dependent on each assignment).

The following theorem shows that LUPFs indeed serve as upper bounds for $\supval(\centerdot)$.

\smallskip
\begin{theorem}\label{thm:lmf}\label{thm:upbound}
If $h$ is an LUPF, then $\supval(\pv)\le h(\pv)-K$ for all valuations $\pv\in\Rset^{|X|}$ such that $\pv\models\phi$.
\end{theorem}

\noindent{\em Proof sketch.} The key ideas of the proof are as follows:
Fix any scheduler $\sigma$ that ensures finite expected termination time.
\begin{compactitem}
\item We first construct a stochastic process based on $h$.
Using the condition (C3) which is non-increasing property we establish that
the stochastic process obtained is a supermartingale
(for definitions of supermartingale see~\cite[Chapter~10]{probabilitycambridge}).
The supermartingale in essence preserve the non-increasing property.

\item Given the supermartingale, we apply Optional Stopping Theorem (OST) (\cite[Chapter~10.10]{probabilitycambridge}),
and use condition (C4) to establish the required boundedness condition of OST,
to arrive at the desired result.
\end{compactitem}
While conditions (C1) and (C2) are not central to the proof, the condition (C1) ensures linearity,
which will be required by our algorithms, and the condition (C2) is the boundedness after termination,
that derives the desired upper bounds (i.e., contribution of the term $K$ comes from condition (C2)).

\smallskip\noindent{\em Formal proof of Theorem~\ref{thm:lmf}.}
In order to formally prove the theorem, we need the fundamental mathematical notions of filtrations, stochastic processes and conditional expectation.

\noindent{\em Filtrations
	.}
A \emph{filtration} of a probability space $(\Omega,\mathcal{F},\probm)$ is an infinite sequence $\{\mathcal{F}_n \}_{n\in\Nset_0}$ of $\sigma$-algebras over $\Omega$ such that $\mathcal{F}_n \subseteq \mathcal{F}_{n+1} \subseteq\mathcal{F}$ for all $n\in\Nset_0$.

\noindent{\em Discrete-Time Stochastic Processes.}
A \emph{discrete-time stochastic process} is a sequence $\Gamma=\{X_n\}_{n\in\Nset_0}$ of random variables where $X_n$'s are all for some probability space (say, $(\Omega,\mathcal{F},\probm)$);
$\Gamma$ is \emph{adapted to} a filtration $\{\mathcal{F}_n\}_{n\in\Nset_0}$ of sub-$\sigma$-algebras of $\mathcal{F}$ if for all $n\in\Nset_0$, $X_n$ is $\mathcal{F}_n$-measurable.

\noindent{\em Conditional Expectation.}
Let $X$ be any random variable for a probability space $(\Omega, \mathcal{F},\probm)$ such that $\expv(|X|)<\infty$.
Then given any $\sigma$-algebra $\mathcal{G}\subseteq\mathcal{F}$, there exists a random variable (for $(\Omega, \mathcal{F},\probm)$), conventionally denoted by $\condexpv{X}{\mathcal{G}}$, such that
\begin{compactitem}
	\item[(E1)] $\condexpv{X}{\mathcal{G}}$ is $\mathcal{G}$-measurable, and
	\item[(E2)] $\expv\left(\left|\condexpv{X}{\mathcal{G}}\right|\right)<\infty$, and
	\item[(E3)] for all $A\in\mathcal{G}$, we have $\int_A \condexpv{X}{\mathcal{G}}\,\mathrm{d}\probm=\int_A {X}\,\mathrm{d}\probm$.
\end{compactitem}
The random variable $\condexpv{X}{\mathcal{G}}$ is called the \emph{conditional expectation} of $X$ given $\mathcal{G}$,
and is unique in the sense that if $Y$ is another random variable satisfying (E1)--(E3), then $\probm(Y=\condexpv{X}{\mathcal{G}})=1$.
See~\cite[Chapter~9]{probabilitycambridge} for more details.

\begin{proof}[Proof of Theorem~\ref{thm:upbound}]
	Fix any scheduler $\sigma$ and initial valuation $\pv$ for our simple while loop.
	Let $T$ be the random variable that measures the number of loop iterations.
	By our assumption, $\expv(T)<\infty$ under $\sigma$.
	Define the following sequences of (vectors of) random variables:
	\begin{compactitem}
		\item $\overline{\pv}_0,\overline{\pv}_1,\dots$ where each $\pv_n$ represents the valuation before the $n$th loop iteration of the while loop (so that $\overline{\pv}_0=\pv$);
		\item $\overline{\rv}_0,\overline{\rv}_1,\dots$ where each $\rv_n$ represents the sampled valuation for the $n$th loop iteration of the while loop;
		\item $\overline{\ell}_0,\overline{\ell}_1,\dots$ where each $\overline{\ell}_n$ represents the nondeterministic choice from the scheduler for the $n$th loop iteration.
	\end{compactitem}
	We also recall the random variables $C_0,C_1,\dots$ where each $C_n$ represents the cost/reward during  the $n$th iteration.
	
	By the execution of the loop, we have:
	\[
	\overline{\pv}_{n+1}=
	\begin{cases}
	F_{\overline{\ell}_n}(\overline{\pv}_n,\overline{\rv}_n) & \mbox{if }  \overline{\pv}_n\models\phi  \\
	\overline{\pv}_n & \mbox{otherwise}
	\end{cases},
	\]
	
	\[
	C_{n}=
	\begin{cases}
	R(\overline{\rv}_n,\overline{\ell}_n) & \mbox{if }  \overline{\pv}_n\models\phi  \\
	0 & \mbox{otherwise}
	\end{cases}\enskip.
	\]
	We also have that $T=\min\{n\mid \overline{\pv}_n\models \neg\phi\}$.
	Then we define the stochastic process $Y_0,Y_1,\dots$ by:
	\[
	\textstyle Y_n:=h(\overline{\pv}_n)+\sum_{m=1}^{n-1} C_m\enskip.
	\]
	We accompany $Y_0,Y_1,\dots$ with the filtration $\mathcal{F}_0,\mathcal{F}_1,\dots$ such that each $\mathcal{F}_n$ is the smallest sigma-algebra that makes all random variables from $\overline{\pv}_0,\dots,\overline{\pv}_n,\overline{\rv}_0,\dots,\overline{\rv}_{n-1}$ and $\overline{\ell}_0,\dots,\overline{\ell}_{n-1}$ measurable. Hence,
	\begin{eqnarray*}
		& &\condexpv{Y_{n+1}}{\mathcal{F}_n}\\
		&=& \condexpv{Y_{n}+h(\overline{\pv}_{n+1})-h(\overline{\pv}_n)+C_{n}}{\mathcal{F}_n}\\
		&=& Y_n+\left(\condexpv{h(\overline{\pv}_{n+1})+C_{n}}{\mathcal{F}_n}-h(\overline{\pv}_n)\right)\\
		&=& Y_n-h(\overline{\pv}_n)+\mathbf{1}_{\overline{\pv}\models\neg\phi}\cdot h(\overline{\pv}_n) +\\
		& &\mathbf{1}_{\overline{\pv}_n\models\phi}\cdot\left[\condexpv{h(F_{\overline{\ell}_n}(\overline{\pv}_n,\overline{\rv}_n))}{\mathcal{F}_n}+\condexpv{R(\overline{\rv}_n,\overline{\ell}_n)}{\mathcal{F}_n}\right]\\
		&=& Y_n-h(\overline{\pv}_n)+\mathbf{1}_{\overline{\pv}_n\models\neg\phi}\cdot h(\overline{\pv}_n)\\
		& &\quad{}+\mathbf{1}_{\overline{\pv}_n\models\phi}\cdot\left[\expv_{\mathbf{u}}\left({h(F_{\overline{\ell}_n}(\overline{\pv}_n,\mathbf{u}))}\right)+\expv_{\mathbf{u}}\left({R(\mathbf{u},\overline{\ell}_n)}\right)\right]\\
		&\le& Y_n \mbox{ (by (C3))}
	\end{eqnarray*}
	where $\mathbf{1}_{\overline{\pv}_n\models\phi}$ is the random variable such that
	\[
	\mathbf{1}_{\overline{\pv}_n\models\phi}=
	\begin{cases}
	1 & \mbox{if }  \overline{\pv}_n\models\phi  \\
	0 & \mbox{otherwise}
	\end{cases}
	\]
	and $\mathbf{1}_{\overline{\pv}_n\models\neg\phi}=1-\mathbf{1}_{\overline{\pv}_n\models\phi}$.
	Hence, $\{Y_n\}_{n\in\Nset_0}$ is a supermartingale. Moreover, we have from (C4) that $|Y_{n+1}-Y_n|\le M+R_{\max}$.
	Thus, by applying Optional Stopping Theorem, we obtain immediately that $\expv(Y_T)\le \expv(Y_0)$.
	By definition,
	\[
	\textstyle Y_T=h(\overline{\pv}_T)+\sum_{m=1}^{T-1} C_m\enskip.
	\]
	It follows from (C2) that $\expv(C_\infty)=\expv(\sum_{m=1}^{T-1} C_m)\le \expv(Y_0)-K=h(\pv)-K$.
	Since the scheduler $\sigma$ is chosen arbitrarily, we obtain that $\mbox{\sl supval}(\pv)\le h(\pv)-K$.
\end{proof}

\subsubsection{Lower bounds}

For the lower bound, we have the following definition:

\smallskip
\begin{definition}[Linear Lower Potential Functions (LLPFs)]
A \emph{linear lower potential function (LLPF)} is a function $h:\Rset^{|X|} \rightarrow\Rset$
that satisfies (C1),(C2),(C4) and in addition (C3') (instead of (C3)) as follows:
\begin{compactitem}
\item[\emph{(C3')}] there exists $\ell\in N$ such that
\[
h(\pv)\le \expv_{\rv}(h(F_\ell(\pv,\rv)))+\expv_{\rv}(R(\rv,\ell))
\]
for all $\pv$ satisfying $\pv\models\phi$.
\end{compactitem}
\end{definition}

Similar to Theorem~\ref{thm:lmf}, we obtain the following result on lower bounds for
$\supval(\centerdot)$.

\smallskip
\begin{theorem}\label{thm:llmf}
If $h$ is an LLPF, then $\supval(\pv)\ge h(\pv)-K'$ for all valuations $\pv\in\Rset^{|X|}$ such that $\pv\models\phi$.
\end{theorem}

\begin{proof}
	This theorem can be proved in the same manner as Theorem~\ref{thm:upbound}.
\end{proof}

%% file: algorithm_arxiv.tex
\section{Computational Results}\label{sect:algorithm}

By Theorem~\ref{thm:lmf} and Theorem~\ref{thm:llmf}, to obtain tight upper and lower bounds
for the SSP problem, we need an algorithm to obtain good LUPFs and LLPFs, respectively.
We present the results for upper and lower bounds separately.

\subsection{Computational Approach for Upper Bound}

The key steps to obtain an algorithmic approach is as follows:
(i)~we first establish a linear template with unknown coefficients  for a LUPF from (C1);
(ii)~then we transform logical conditions (C2)--(C4) equivalently into inclusion assertions between polyhedrons;
(iii)~next we transform inclusion assertions into linear inequalities through Farkas' Lemma;
(iv)~finally we solve the linear inequalities through linear programming, where the solution for unknown coefficients in the template synthesizes a concrete LUPF that serves as an upper bound for the $\sup$-value.
Below we recall the well-known Farkas' Lemma.

\smallskip
\begin{theorem}[Farkas' Lemma~\cite{FarkasLemma}]\label{thm:farkas}
Let $\mathbf{A}\in\mathbb{R}^{m\times n}$, $\mathbf{b}\in\mathbb{R}^m$, $\mathbf{c}\in\mathbb{R}^{n}$ and $d\in\mathbb{R}$.
Suppose that $\{\mathbf{x}\in\mathbb{R}^n\mid \mathbf{A}\mathbf{x}\le \mathbf{b}\}\ne\emptyset$.
Then
\[
\{\mathbf{x}\in\mathbb{R}^n\mid \mathbf{A}\mathbf{x}\le \mathbf{b}\}\subseteq \{\mathbf{x}\in\mathbb{R}^n\mid \mathbf{c}^{\mathrm{T}}\mathbf{x}\le d\}
\]
iff there exists $\mathbf{y}\in\mathbb{R}^m$ such that $\mathbf{y}\ge \mathbf{0}$, $\mathbf{A}^\mathrm{T}\mathbf{y}=\mathbf{c}$ and $\mathbf{b}^{\mathrm{T}}\mathbf{y}\le d$.
\end{theorem}
Intuitively, Farkas' Lemma transforms the inclusion problem of a nonempty polyhedron within a halfspace into a feasibility problem of a system of linear inequalities.
As a result, one can decide the inclusion problem in polynomial time through linear programming.

\smallskip\noindent{\bf The Algorithm $\mathsf{UpperBound}$.}
Consider as input a succinct MDP in the form (\ref{eq:swl}).
\begin{compactenum}
\item {\em Template.} The algorithm sets up a column vector $\mathbf{a}$ of $|X|$ fresh variables and a fresh scalar variable $b$
such that the template for an LUPF $h$ is
$h(\pv)=
\mathbf{a}^{\mathrm{T}}\cdot\pv+b$.
\item {\em Constraints on $\mathbf{a}$ and $b$.} The algorithm first encodes the condition (C2) for the template $h$ as the inclusion assertion
\begin{align*}
&\{(\pv,\rv)\mid \pv\models\phi, F_\ell(\pv,\rv)\models\neg\phi\}\\
&\qquad\subseteq \{(\pv,\rv)\mid K\le \mathbf{a}^{\mathrm{T}}\cdot F_\ell(\pv,\rv)+ b\le K'\}
\end{align*}
parameterized with $\mathbf{a},b, K,K'$ for every $\ell\in N$,
where $K,K'$ are fresh unknown constants.
Then for every $\ell\in N$, the algorithm encodes (C3) as
\[
\{\pv\mid \pv\models\phi\}\subseteq \{\pv\mid \mathbf{c}_\ell^{\mathrm{T}}\cdot\pv\le d_\ell\}
\]
where $\mathbf{c}_\ell, d_\ell$ are unique linear combinations of unknown coefficients $\mathbf{a},b$ satisfying that $\mathbf{c}_\ell^{\mathrm{T}}\cdot\pv\le d_\ell$ is equivalent to $h(\pv)\ge \expv_{\rv}(h(F_\ell(\pv,\rv)))+\expv_{\rv}(R(\rv,\ell))$.
Finally, the algorithm encodes (C4) as inclusion assertions with a fresh unknown constant $M$ using similar transformations.
All the inclusion assertions (with parameters $\mathbf{a},b,K,K',M$) are grouped \emph{conjunctively} so that these inclusions should all hold.
\item {\em Applying Farkas' Lemma.} The algorithm applies Farkas' Lemma to all the inclusion assertions generated in the previous step and obtains a system of linear inequalities
involving the parameters $\mathbf{a},b,K,K', M$.
\item {\em Constraint Solving.} The algorithm calls a linear programming solver on the linear program consisting of the system of linear inequalities generated in the previous step and the minimizing objective function $\mathbf{a}^\mathrm{T}\cdot \pv_0+b-K$ where $\pv_0$ is an initial valuation for program variables.
\end{compactenum}

\noindent{\em Correctness and running time.}
The above algorithm obtains concrete values for $\mathbf{a},b,K,K',M$ and leads to a concrete LUPF
$h$. The correctness that $\mathbf{a}^\mathrm{T}\cdot \pv+b-K$ is an upper bound for the $\sup$-value follows from
Theorem~\ref{thm:lmf}.
The main optimization solution required by the algorithm is linear programming, and thus our algorithm runs
in polynomial time in the size of the input succinct MDP.

\smallskip
\begin{theorem}
Given a succinct MDP and the SSP problem, the best linear upper bound (wrt an initial valuation) on the sup-value can be computed in
polynomial-time in the implicit description of the MDP.
\end{theorem}


\smallskip
\begin{example} \label{upexample}
Consider the Gambler's Ruin example (from Section~\ref{sect:runningexamples}, Figure~\ref{fig:gambler}).

Let $h: \mathbb{R} \rightarrow \mathbb{R}$ be an LUPF for this example, we have:
$$
\begin{matrix*}[l]
(C1) & \exists \lambda_1, \lambda_2 \in \mathbb{R} ~~~\forall x \in \mathbb{R}~~~ h(x) = \lambda_1 x + \lambda_2\\
(C2) & \exists K, K' \in \mathbb{R} ~~~ \forall x \in \left[1, 2\right) ~~~ K \leq h(x) \leq K' \\
(C3) & \forall x \in \left[1, \infty\right) ~ h(x) \geq 0.4 \cdot \left( 1 \!+\! h(x\!+\!1) \right) \!+\! 0.6 \cdot h(x\!-\!1)\\
(C3) & \forall x \in \left[1, \infty\right) ~ h(x) \geq 0.3 \cdot \left( 1\!+\! h(x\!+\!1) \right) \!+\!0.7 \cdot h(x\!-\!1)\\
\end{matrix*}
$$\vspace{-3mm} $$
\begin{matrix*}[l]
(C4) & \exists M \in \left[0, \infty\right) \forall x \in \left[1, \infty\right) & \vert h(x) - h(x-1) \vert \leq M\\
(C4) & ~~~~~~~~~~~~~~~~~~~~~~~~~~~~~~~~~~~ \textbf{and} & \vert h(x) - h(x+1) \vert \leq M\\
\end{matrix*}
$$
Note that for condition (C2) we need to quantify over $x \in  \left[1, 2\right)$, as if $x$ is not in this range, then the loop does not terminate in the next iteration.
Given condition (C1), the two (C4) conditions are equivalent to $M \geq \lambda_1$.
Also, the (C2) condition is equivalent to $K \leq \min\{h(1), h(2)\}$ and $K' \geq \max\{h(1), h(2)\}$ or more precisely $K \leq \lambda_1 + \lambda_2$, $K \leq 2 \lambda_1 + \lambda_2$, $K' \geq \lambda_1 + \lambda_2$ and $K' \geq 2 \lambda_1 + \lambda_2$. By expanding the occurrences of $h$ in the first (C3) condition and simplifying, we get \mbox{$\forall x \in \left[1, \infty \right)~~ 0 \geq0.4 - 0.2 \lambda_1$} and we can drop the quantification given that $x$ does not appear. Similarly, the second (C3) condition is equivalent to $0 \geq 0.3 - 0.4 \lambda_1$. In our method, such equivalences are automatically obtained by applying Farkas' Lemma rather than manual inspection of the inequalities.
Now that all necessary conditions are replaced by equivalent linear inequalities, we can solve the linear program to find an LUPF. An optimal answer (with minimal $\lambda_1$) is the following: $\lambda_1 = M = 2, \lambda_2 = K = 0, K' = 4$. Therefore by Theorem \ref{thm:upbound}, we have $\supval(x_0)\le 2x_0$ for all initial valuations $x_0$ that satisfy the loop guard. \qed
\end{example}

\smallskip
\begin{remark}\label{remark:basic-ext}
Note that our approach is applicable to succinct MDPs with integer as well as real-valued variables,
(i.e., the underlying state-space of the MDP is infinite).
Even when we consider integer variables, since $h$ gives upper bounds, the reduction
is to linear programming, rather than integer linear programming.
Note that our approach only depends on expectation of sampling variables, and 
thus applicable even to continuous sampling variables with same expectation,
e.g., our results apply uniformly to Example~\ref{examp:gr} and Example~\ref{examp:grcts}, 
given that the sampling variable $r$ has same expectation.
\end{remark}

\subsection{Computational Approach for Lower Bound}

The algorithm for lower bound is similar to the upper bound, however, there are some subtle and
key differences.
An important difference is that while in Step~2 of the algorithm for upper bound, there is a
{\em conjunction} of contraints, for the lower bound problem it requires a {\em disjunction}.
This has two important implications: first, we need to consider a generalization of Farkas' lemma
and in this case we use Motzkin's Transposition Theorem (which extends Farkas' Lemma with strict inequalities)
and second, instead of linear programming we require quadratic programming.

\smallskip
\begin{theorem}[Motzkin's Transposition Theorem~\cite{Motzkin}]
Let $\mathbf{A}\in\mathbb{R}^{m\times n}$, $\mathbf{B}\in\mathbb{R}^{k\times n}$ and
$\mathbf{b}\in\mathbb{R}^m,\mathbf{c}\in\mathbb{R}^k$.
Assume that $\{\mathbf{x}\in\mathbb{R}^n\mid \mathbf{A}\mathbf{x}\le \mathbf{b}\}\ne\emptyset$.
Then
\[
\{\mathbf{x}\in\mathbb{R}^n\mid \mathbf{A}\mathbf{x}\le \mathbf{b}\}\cap\{\mathbf{x}\in\mathbb{R}^n\mid \mathbf{B}\mathbf{x}<\mathbf{c}\}=\emptyset
\]
iff there exist $\mathbf{y}\in\mathbb{R}^m$ and $\mathbf{z}\in\mathbb{R}^k$ such that $\mathbf{y},\mathbf{z}\ge\mathbf{0}$, $\mathbf{1}^\mathrm{T}\cdot\mathbf{z}>0$,
$\mathbf{A}^{\mathrm{T}}\mathbf{y}+\mathbf{B}^{\mathrm{T}}\mathbf{z}=\mathbf{0}$ and $\mathbf{b}^{\mathrm{T}}\mathbf{y}+\mathbf{c}^{\mathrm{T}}\mathbf{z}\le 0$.
\end{theorem}

\noindent{\bf The Algorithm $\mathsf{LowerBound}$.}
The algorithm here is similar to $\mathsf{UpperBound}$ described previously.
For the sake of brevity, we explain the differences only.

\begin{compactenum}
\item {\em Template $h$.} Same as in the Algorithm $\mathsf{UpperBound}$.
\item {\em Constraints on $\mathbf{a}$ and $b$.} The algorithm first encodes (C2) and (C4) as inclusion assertions in the same way as in $\mathsf{UpperBound}$ and transforms them into linear inequalities over $\mathbf{a},b,K,K',M$ through Farkas' Lemma.
Then the algorithm transforms (C3') equivalently into the inclusion assertion
\[
\textstyle\{\pv\mid \pv\models\phi\}\subseteq \bigcup_{\ell\in N}\{\pv\mid \mathbf{c}_\ell^{\mathrm{T}}\cdot\pv\le d_\ell\}
\]
where $\mathbf{c}_\ell, d_\ell$ are determined in the same way as in $\mathsf{UpperBound}$.
Furthermore, this inclusion assertion is equivalently written as
\[
\textstyle\{\pv\mid \pv\models\phi\}\cap  \bigcap_{\ell\in N}\{\pv\mid \mathbf{c}_\ell^{\mathrm{T}}\cdot\pv> d_\ell\} = \emptyset
\]
and then transformed into a system of quadratic inequalities over $\mathbf{a}$ and $b$ through Motzkin's Transposition Theorem. The system may involve quadratic inequalities since $\mathbf{c}_\ell$ contains the unknown parameters $\mathbf{a}$ and $b$.

\item {\em Constraint Solving.} The algorithm calls a nonlinear-programming solver on the system of linear and quadratic inequalities generated in the previous step with the maximizing objective
function $\mathbf{a}^\mathrm{T}\cdot \pv_0+b-K'$ where $\pv_0$ is an appropriate initial valuation.
\end{compactenum}

\smallskip\noindent{\em Correctness and optimization problem.}
As in the upper bound algorithm, once $\mathbf{a},b,K,K',M$ are found, we obtain an concrete lower bound $\mathbf{a}^\mathrm{T}\cdot \pv+b-K'$
for the $\sup$-value from Theorem~\ref{thm:llmf},
establishing the correctness of our algorithm.
The reduction leads to a quadratic optimization problem of polynomial size in the size of the succinct MDP,
implying the following result.

\smallskip
\begin{theorem}
Given a succinct MDP and the SSP problem, the best linear lower  bound (wrt an initial valuation) on the sup-value can be computed via
a polynomial (in the implicit description of the MDP) reduction to quadratic programming.
\end{theorem}


\smallskip
\begin{example}
Let $h: \mathbb{R} \rightarrow \mathbb{R}$ be an LLPF for the Gambler's Ruin example (Figure~\ref{fig:gambler} in Section~\ref{sect:runningexamples}). The conditions of the form (C1), (C2) and (C4) are exactly the same as in Example~\ref{upexample}. In addition, $h$ must also satisfy the following condition:
$$
\begin{matrix*}[l]
(C3') & \forall x \in \left[1, \infty\right)\\ & h(x) \leq 0.4 \cdot (1 + h(x+1)) + 0.6 \cdot h(x-1)\\
& ~~~~~~~~~~~~~~~~~~~~~~~~~~~~~~~~\textbf{or}\\
& h(x) \leq 0.3 \cdot (1 + h(x+1)) + 0.7 \cdot h(x-1)
\end{matrix*}
$$
Expanding the occurrences of $h$ in the condition above using $h(x) = \lambda_1 x + \lambda_2$, and discarding the quantification, we obtain the following equivalent disjunctive system of inequalities: $0 \leq 0.4 - 0.2 \lambda_1$ or $0 \leq 0.3 - 0.4 \lambda_1$. This system is obviously equivalent to $\lambda_1 \leq 2$. Note that in general we have disjunction of linear inequalities, which require quadratic programming. As explained previously, such equivalences are automatically obtained by our algorithm using Motzkin's transposition theorem.

Adding the equivalent linear forms of conditions (C1), (C2) and (C4) as in Example~\ref{upexample} and considering the resulting linear program with the objective of maximizing $\lambda_1$ leads to the following solution: $\lambda_1 = M = 2, \lambda_2 = -2, K=0, K' = 2$. Therefore, by Theorem~\ref{thm:llmf}, $\supval(x_0)\ge 2x_0$ for every initial valuation $x_0$ that satisfies the loop guard. Given that this is the exact same upper bound we found in Example~\ref{upexample}, the bound is tight.\qed
\end{example}

\subsection{Remarks}

\medskip
\begin{remark}[Sampling Distributions]
For simplicity, we only considered discrete and uniform sampling distributions in our examples in this paper. However, as mentioned in Remark~\ref{remark:basic-ext}, 
since we only use the mean of random variables, our approach extends to other sampling distributions with known means.
\end{remark}

\medskip
\begin{remark}[Dependence of Rewards on Program Variables]
Our proof (of Theorem~\ref{thm:upbound} that is the basis of all results) depends on the Optional Stopping Theorem that requires bounded difference in all steps. 
We considered that the rewards depend only on sampling variables and non-determinism. This assumption is sufficient, but not always necessary, to ensure bounded difference. 
Our approach is also applicable if the rewards depend on program variables, provided that they remain bounded.
\end{remark}

\medskip
\begin{remark}
Our approach computes the best \emph{linear} upper and lower bounds wrt the initial valuations. However, a succinct MDP might collect logarithmic reward. For example, the succinct MDP shown in Figure~\ref{fig:logarithmic} halves the variable $x$ until it becomes less than or equal to $1$, and has a unit reward at each step. Hence, it collects $\lg x$ reward in total. In such cases, the obtained bounds, which are linear, can be arbitrarily bad. 
	
\begin{figure}[!hb]
		\begin{lstlisting}[mathescape]
while $x \ge 1$ do
    ? $x := x / 2$  reward=$1$ 
		\end{lstlisting}
		
		\caption{A succinct MDP with logarithmic total reward}
		\label{fig:logarithmic}
		
	\end{figure}
\end{remark}

\smallskip\noindent{\em Future work.}
While in this work we consider succinct MDPs, which are special case of factored MDPs, 
extending our approach to factored MDPs with linear dependency on the variables, 
but without restriction on nesting structure of while-loops, 
is an interesting direction of future work.

%% file: cases_arxiv.tex
\section{Case Studies and Experiments}

We present more case studies and our experimental results.

\subsection{Additional Examples}

We consider several other classical problems in probabilistic planning that can be described as succinct MDPs.
We provide examples of Robot Planning and two variants of Roulette as typically played in casinos.

\noindent \textbf{Two-dimensional Robot Planning.} Consider a robot that is placed on an initial point $(x_0, y_0)$ of a two-dimensional grid. A player controls the robot and at each step, can order it to move one unit in either direction (left, right, up, down). However, the robot is not perfect. It follows the order with probability $p < 1$ and ignores it and goes to the left with probability $1 - p$. The process ends when the robot leaves the $x \geq y$ half-plane and the player's objective is to keep the robot in this half-plane. The player gets a reward of $1$ each time the robot moves. The half-plane was chosen arbitrarily for demonstration of our approach. Our method can handle any half-plane and starting point.

\noindent \textbf{Multi-robot Planning.} Our approach can handle as many variables as necessary and is only polynomially dependent on the succinct representation of the MDP. To demonstrate this, we consider a scenario similar to the previous case, in which there are now two robots $r_1$ and $r_2$ starting at positions $(x_0, y_0)$ and $(x'_0, y'_0)$. The robot $r_1$ follows the orders with probability $p_1$ and malfunctions and goes right with probability $1-p_1$. Similarly, $r_2$ follows the commands with probability $p_2$ and goes left with probability $1-p_2$. The player's goal is to keep $r_1$ to the left of $r_2$,~i.e. to keep the robots in the four-dimensional half-space $x \leq x'$.
The process ends when the robots leave this half-space and the player gets a reward of $1$ per step as long as the process has not terminated. 

\noindent \textbf{Mini-roulette.} We model Mini-roulette which is a popular casino game based on a 13-slot wheel. A player starts the game with $x_0$ chips. She needs one chip to make a bet and she bets as long as she has chips. If she loses a bet, the chip will not be returned, but a winning bet will not consume the chip and results in a specific amount of (monetary) reward, and possibly even more chips.
The following types of bets can be placed at each round:
(i)~\emph{Even money bets:} In these bets, 6 specific slots are chosen. Then the ball is rolled and the player wins the bet if it lands in one of the 6 slots. So the player has a winning probability of $6/13$. Winning them gives a reward of one unit and one extra chip.
(ii)~\emph{2-to-1 bets:} These bets correspond to 4 slots and winning them gives a reward of 2 and 2 extra chips.
(iii,iv,v)~\emph{3-to-1, 5-to-1 and 11-to-1 bets:} These are defined similarly and have winning probabilities of $3/13, 2/13$ and $1/13$ respectively.

	
\noindent \textbf{American Roulette.} This game is similar to Mini-roulette in all aspects, except that there are more complicated
	types of bets available to the player and that the wheel has 38 slots. The player can now have half-chips and can bet as long as he has at least one chip remaining. A bet can lead to one of three outcomes. A definite loss, a partial loss or a win. A
	definite loss costs one chip and a partial loss half a chip. A win leads to some reward
	and more chips. Table \ref{table:americanRoulette} summarizes the types of bets available in the game.
	
	\begin{table}[!hb]
		\centering
		\resizebox{.7\columnwidth}{!}{%
			
		\begin{tabular}{|C{1.1cm}|C{1.6cm}|C{1.6cm}|C{1.1cm}|C{1.1cm}|}
			\hline
			Type & Winning Probability & Partial Loss Probability & Winning Reward & Winning Tokens  \\
			\hline
			\hline
			35-to-1 & 1/38                & 0                        & 35              & 35              \\
			17-to-1 & 1/19                & 0                        & 17              & 17              \\
			11-to-1 & 3/38                & 0                        & 11              & 11              \\
			8-to-1  & 2/19                & 0                        & 8              & 8               \\
			6-to-1  & 5/38                & 0                        & 6             & 6               \\
			5-to-1  & 3/19                & 0                        & 5              & 5               \\
			2-to-1  & 6/19                & 1/19                     & 2             & 2               \\
			Even  & 9/19                & 1/19                     & 1              & 1     \\
			\hline
		\end{tabular}}
	\caption{Types of bets available in American Roulette.}
	\label{table:americanRoulette}
	\end{table}

%% file: experiments_arxiv.tex
\subsection{Experimental Results}

\begin{table}
	\centering
	\resizebox{\columnwidth}{!}{
	\begin{tabular}{|C{2cm}|C{1.4cm}|C{2.1cm}|C{1.9cm}|C{1.1cm}|}
		\hline
		MDP & Parameters & Upper bound & Lower bound & Time   \\
		\hline \hline
		Gambler's Ruin & $p_1 = 0.4$ $p_2 = 0.3$ &  $2x$ & $2x$ & $153$ ms\\
		\hline
		2D Robot Planning & $p = 0.4$ & $5x - 5y$ & $5x - 5y$ & $251$ ms \\
		\hline
		Multi-robot Planning & $p_1 = 0.4$ $p_2 = 0.4$ & $2.5 x - 2.5 y + 5$ & $2.5 x - 2.5 y$ & $758$ ms \\
		\hline
		Mini-roulette & --- & $11x$ & $11x$ & $320$ ms \\
		\hline
		American Roulette & --- & $24x$ & $24x$ & $425$ ms
		\\
		\hline

	\end{tabular}
}
	\caption{Experimental Results}

	\label{table:exps}
\end{table}

We implemented our approach in Java and obtained experimental results on all examples mentioned previously. The results are summarized in Table~\ref{table:exps}, where  ``Parameters'' shows concrete parameters for our examples (where for the last two examples there is no parameter), ``Upper bound'' (resp. ``Lower bound'') presents the LUPFs (resp. LLPFs) obtained through our approach, and finally ``Time'' shows the running time in milliseconds.
The reported upper bounds on $\sup$-values are the results of $h(\pv) - K$ as in Theorem~\ref{thm:upbound}. Similarly, the reported lower bounds on $\sup$-values are obtained from $h(\pv)-K'$ as in Theorem~\ref{thm:llmf}.
All initial valuations lead to the same results in our experiments.
Finally, our approach is not sensitive to parameters as varying parameters will only change coefficients of our LUPFs/LLPFs. 

\noindent{\em Runtime and Platform.} Note that our approach is extremely efficient and can handle all these MDPs, even those with large succinct representation such as the American Roulette MDP, in less than a second. The results were obtained on a machine with Intel Core i5-2520M dual-core processor (2.5GHz), running Ubuntu Linux 16.04.3 LTS. We used lpsolve \cite{berkelaar2004lpsolve}, JavaILP \cite{javailp} and JOptimizer \cite{joptimizer} for solving linear and quadratic optimization tasks.

\noindent{\em Significance of our results.}
First, observe that in most experimental results the upper and lower bounds are tight.
Thus our approach provides precise bounds on the SSP problem for several classical examples.
Second, our results apply to infinite-state MDPs: in all the above examples, we consider infinite-state
MDPs, where algorithmic approaches for factored MDPs do not work.
Finally, in the above examples, instead of infinite-state MDPs if we consider large finite-state MDP
variants (e.g., bounding the variable $x$ in Gambler's Ruin with a large domain), then
as the MDP becomes larger, the SSP value of the finite-state MDP approaches the infinite-state value.
Since we provide tight bounds on the SSP value for this infinite-state limit, our approach provides
efficient approximation even for large finite-state MDP variants.

%% file: related_arxiv.tex
\section{Related Works} \label{sec:related}

\noindent{\em MDPs.}
MDPs have been studied quite widely and deeply in the AI literature~\cite{Sigaud:2010:MDP:1841781,Dean:1997:MRT:2074226.2074241,Singh1994284,WILLIAMS2007393,DBLP:conf/aaai/PoupartMPKGB15,DBLP:conf/aaai/GilbertWX17,DBLP:conf/ijcai/TopinHSWdM15,DBLP:conf/ijcai/PerraultB17,DBLP:conf/uai/BoutilierL16,Ferns:2004:MFM:1036843.1036863};
and factored MDPs have also been considered as an efficient algorithmic approach~\cite{DBLP:journals/jair/GuestrinKPV03}.
In this line of work, we introduce succinct MDPs and efficient algorithms for them which
are applicable to several problems in AI.

\noindent{\em PPDDL and RDDL.}
There are a variety of languages for specifying MDPs and especially factored MDPs. Two of the most commonly used are PPDDL~\cite{younes2004ppddl1} and RDDL~\cite{sanner2010relational}. These languages are general languages that capture all factored MDPs. Instead, we consider succinct MDPs where guards and assignments are linear expressions and we do not consider nested while-loops, and thus our language is simpler than PPDDL and RDDL.

\noindent{\em Programming language results.}
Besides the AI community, research in programming languages also considers probabilistic
programs and algorithmic approaches~\cite{DBLP:conf/cav/ChakarovS13,chatterjee2016};
but the main focus is termination with probability~1 or in finite time, whereas
we consider the SSP problem  and compute precise bounds for it. 
While both approaches use the theory of martingales as the underlying mathematical tool, 
the key differences are as follows:
\begin{itemize}
	
\item \emph{Problem difference:} The work of~\cite{chatterjee2016} considers the number of steps to termination.
There is no notion of rewards or stochastic shortest path (SSP). In contrast, we consider rewards and SSP. 
In particular, we have negative rewards that cannot be modeled by the notion of steps.
	
\item \emph{Result difference:} \cite{chatterjee2016} considers the qualitative question of whether expected termination time is finite or not, and then applies Azuma's inequality to martingales for concentration bounds. 
In contrast, we present upper and lower bounds on expected SSP. Thus our results provide quantitative 
(rather than qualitative) bounds for expected SSP that significantly generalize expected termination time. 
However, our results are applicable to a more restricted class of programs.
	
\item \emph{Proof-technique difference:} \cite{chatterjee2016} considers qualitative expected termination time 
characterization, and the main mathematical tool is martingale convergence that does not handle negative rewards. 
In contrast, we present quantitative bounds for SSP and our mathematical tool is Optional Stopping Theorem.
\end{itemize}

\noindent{\em Comparison with \cite{hansen}.} 
This work provides convergence tests for heuristic search value-iteration algorithms. 
While both approaches provide bounds for SSPs, the main differences are as follows: 
(i)~our approach can handle negative costs whereas~\cite{hansen} can handle only positive costs;
(ii)~our results are on the implicit representation of MDPs, while~\cite{hansen} evaluates parts of the explicit MDP; and  
(iii)~our approach presents polynomial reductions to optimization problems and is 
not dependent on value-iteration.

%% file: conclusion_arxiv.tex
\section{Conclusion}

We consider succinct MDPs, which can model several classical examples from the AI literature,
and present algorithmic approaches for the SSP problem on them.
There are several interesting directions for future work.
The first direction would be to consider other algorithmic approaches for succinct MDPs.
In our work, we consider linear templates for efficient algorithmic analysis.
Generalization of our approach to more general templates is another interesting direction for future work.
Finally, whether our approach can be extended to other class of MDPs is also an interesting problem to investigate.